\documentclass[aps,twocolumn,10pt]{revtex4}
\usepackage{amsfonts}
\usepackage{amsmath}
\usepackage{amssymb}
\usepackage{graphicx}
\usepackage{epsfig}

\begin{document}

\title{Detecting multi-atomic composite states in optical lattices}
\author{Anatoly Kuklov$^{1}$ and  Henning Moritz$^{2}$}
\affiliation{$^{1}$Department of Physics,CSI, CUNY - Staten Island, New York, NY 10314\\
$^{2}$Institute of Quantum Electronics, ETH Zurich, 8093 Zurich,
Switzerland}

\begin{abstract}

We propose and discuss methods for detecting quasi-molecular complexes which
are expected to form in strongly interacting optical lattice systems.
Particular emphasis is placed on the detection of composite fermions forming
in Bose-Fermi mixtures.  We argue that, as an indirect indication of the
composite fermions and a generic consequence of strong interactions,
periodic correlations must appear in the atom shot noise of bosonic
 absorption images, similar to the bosonic Mott insulator
[S. F\"olling, et al., Nature {\bf 434}, 481 (2005)]. The composites can
also be detected directly and their quasi-momentum distribution measured.
This method -- an extension of the technique of noise correlation
interferometry [E. Altman et al., Phys. Rev. A {\bf 79}, 013603 (2004)] --
relies on measuring higher order correlations between the bosonic and
fermionic shot noise in the absorption images. However, it fails for
complexes consisting of more than three atoms.
\end{abstract}

\maketitle

\section{Introduction}\label{intro}

Atomic mixtures in optical lattices represent a novel laboratory system for
the study of ultracold matter. They are intrinsically clean and feature a
unique amount of tunability. Therefore they offer outstanding opportunities
to investigate fundamental many-body systems exhibiting some very rich
physics. A fascinating example is the formation of composite states, which
are induced by the lattice and the interaction. Such states have recently
been detected experimentally \cite{exp1,exp2,exp4,Volz2006}.

Two-component mixtures of atoms of, say, sorts A and B provide the simplest
example of such objects which can be viewed as diatomic AB quasi-molecules
occurring if the A-B interaction is strong enough. It is interesting to note
that the formation of composites is possible not only for attractive
\cite{Kagan,Efremov2002} but also for repulsive interaction \cite{SCF}. In the last case
the pairing occurs between an atom of sort, say, A and a hole $\overline{\rm B}$
of the sort B. Moreover,
multi-atomic complexes of the type AB$_n$ with $n>1$ can form as well 
\cite{Kagan_trio,Erich,Lewenstein2004}.  Even in strongly non-equilibrium situations
quasi-bound atomic states can exist if  the repulsive energy greatly exceeds
a typical single particle bandwidth. This mechanism was first introduced for
the point-like defects in solid He4 \cite{Andreev} and realized recently in
the optical lattice \cite{Zoller}.

In this paper particular emphasis will be put on generic schemes for
detecting the composites in the lattice Bose-Fermi mixtures since such mixtures have
recently been studied experimentally
\cite{ETH1,Ospelkaus2006}. We will be referring to the composites of the
type FB$_n$ or F$\overline{\rm B}_n$, $n=1,2,3,...$ where  F stands for a
fermion and B ($\overline{\rm B}$) denotes a boson (bosonic hole). However,
all the conclusions can be easily extended to any other type of a composite.

The quasi-molecular complexes are induced by the lattice and consequently
 vanish in most cases once the lattice potential is removed. Since the
standard method of detection in ultracold gases, namely time of flight
imaging, entails the elimination of all trapping potentials, the detection
of composites with this technique is challenging. In the example of
Bose-Fermi mixtures, the formation of such complexes --- composite
fermions--- would lead to the destruction of the Fermi surface of the
original fermions. Therefore the disappearance of the corresponding feature
should be observable in the absorption image. However, this disappearance
would by no means constitute a definite proof of the formation of composites
since a variety of phenomena, e.g. heating, could also lead to the same
observation.

In this letter we propose a method how such composites can be detected
directly. The method is  similar to the Hanbury Brown and Twiss 
noise correlation interferometry \cite{HBT} applied to thermal sources of
cold atoms \cite{ions} and proposed in ref.\cite{Demler2} for revealing
non-trivial many-body states in ultracold atoms. Such a method has been
successfully implemented in ref. \cite{Bloch_noise,Aspect,Spielman2006} for
bosonic and in ref.\cite{Greiner2005,Bloch_fermi} for fermionic systems.

Noise correlation spectroscopy relies on the fact that, while the momentum
distribution itself shows no particular features, the atomic shot noise is
correlated. In order to introduce the extended noise correlation method with
a simple example, consider a model system with a single composite
(quasi-molecule) consisting of one boson and one fermion. The quasi-momentum
of the center of mass of the composite is denoted by \textbf{Q}. When the
lattice is switched off, the pair breaks and the fermionic and bosonic
wavepackets expand independently from each other. The fermionic or the
bosonic momenta distributions $n_F(\bf q_1)$ and $n_B(\bf q_2)$,
respectively, by themselves, which can be measured after expansion, do not
reveal any low energy structure in their individual momenta $\bf{q}_1,\,
\bf{q}_2$ -- they feature almost uniform backgrounds because binding of both
atoms entails a wide range of their relative momenta. However, momentum
conservation requires the momenta to add up to \textbf{Q}. Therefore, if in
an individual measurement, the fermion is found to have $\bf q_1$, the boson
must have $\bf q_2=\bf Q-\bf q_1$ and the correlation between the two
enables one to measure the quasi-momentum of the composite.

In the more general case with more than one pair in the system, the
correlations between the bosonic and fermionic momentum distributions, which
allow one to determine the quasi-momenta of the pairs, will only be
observable in the noise $\delta n_{B,F}=n_{B,F} - \langle n_{B,F}\rangle$ of
the measured momentum distributions.  This is due to the fact that, besides
the correlated pair events, there are also events due to the components
originating from different pairs. The key point is that the momentum
distribution of the (centers of mass of the) FB composites is reproduced by
the noise correlator in the far field of free expansion
\begin{equation}
I_{CF}(\textbf{Q})=\int d{\bf X}_2\langle \delta n_F(\textbf{q}_1) \delta
n_B(\textbf{q}_2)\rangle, \label{I_2}
\end{equation}
with $\textbf{Q}=\textbf{q}_1 + \textbf{q}_2$ fixed, and ${\bf q}_1=m_F{\bf
X}_1/t=\textbf{Q}-{\bf q}_2$, ${\bf q}_2=m_B{\bf
X}_2/t$  being the dependent and independent variables, respectively.
Here $m_F,\, m_B$ are the masses of fermionic and
bosonic atoms, respectively, and ${\bf X}_1$ (${\bf X}_2$) denotes the
position in the far field where a fermion (boson) is detected. We point out
that (\ref{I_2}) is essentially the Fermi distribution function
$\tilde{\rho}_{CF}({\bf Q})$ of the composite fermions (
$\tilde{\rho}_{CF}({\bf Q})\approx 1$ for $|{\bf Q}|\leq Q_{CF}$ and
$\tilde{\rho}_{CF}({\bf Q})\approx 0$ for $|{\bf Q}|\geq Q_{CF}$, where
$Q_{CF}$ stands for the Fermi surface wavevector of the composites). 
We also show that in the repulsive Bose-Fermi mixture, the role of the
center of mass momentum $\bf Q$ is played by $\textbf{Q}_-=\textbf{q}_1 -
\textbf{q}_2$ similar to the case of the repulsive two-component bosonic mixture
\cite{Demler2}. These
properties of the correlator (\ref{I_2})
will be discussed in detail in the appendix \ref{strong} and extended
to a general case of FB$_n$ and F$\overline{\rm B}_n$ complexes in section \ref{com}.

Detection of the Fermi edge in $I_{CF}(\textbf{Q})$ may require quite low temperatures $T$
determined by the Fermi energy $E_{CF}$ of the composites. However, since a typical binding energy $E_b$ of the composites
can be much larger than $E_{CF}$,
 a signature of the composites will be seen as long as $ T< E_b$ (for $T>E_b$,
the correlator (\ref{I_2}) is essentially zero).
Indeed, the very fact that $I_{CF}(\textbf{Q})$
is finite and proportional to a typical density of the original fermions or bosons
is a direct indication of the presence of the composites. Accordingly, as $T$ is lowered
below $E_{CF}$, the formation of the Fermi edge in $I_{CF}(\textbf{Q})$  will be observed.

In principle, detection of multi-atomic complexes
--- quasi-molecules with the number of the atomic constituents exceeding two
--- can be done by extending the noise-correlator method, as explained in
Sec.\ref{com}. However, as will be discussed in Sec.\ref{noise}, such an
extension to higher order correlators can lead to a dramatic increase of
uncorrelated noise. For this case, we propose an indirect method to observe
the complexes (Sec.\ref{indirect}). In this method, the presence of
complexes of the type FB$_n$ is revealed by imaging the density-density
noise correlations of the original bosons, if their number is
incommensurate with the number of the lattice sites. Then the resulting
image is similar to the one observed in a bosonic Mott insulator
\cite{Bloch_noise,Spielman2006}. Such a feature (at zero
temperature) turns out to be a consequence of a strongly interacting
many-body ground state in which individual momenta of bosons are
poorly defined and, therefore, strongly fluctuate.

\section{Description and direct detection of composites in optical lattices}

\subsection{The nature of the composite pairs}
Here we will give a brief account of the nature of equilibrium
quasi-molecular states. In general, quasi-molecular (composite) states form
when the binding energy $E_b$ exceeds the temperature $T$ as well as the gain in
kinetic energy the constituents would encounter by delocalizing from each
other. For attractive interactions between bosons and fermions, composites
states of the type FB are formed. For repulsive interaction, on the other
hand, the quasi-molecules represent bound states of one fermion F and a
bosonic hole $\overline{\rm B}$ \cite{SCF} or, similarly, one fermionic hole
$\overline{\rm F}$ and a boson {\rm B}. Binding of one fermion with more
than one boson or bosonic hole is also possible if the number of bosons per
site is large enough and the magnitude of the Bose-Fermi interaction
$|U_{BF}|$ substantially exceeds the repulsion between bosons $U_{BB}$
\cite{Lewenstein2004}. In this event the composites FB$_n$ (for $U_{BF} <0$)
or F$\overline{\rm B}_n$ (for $U_{BF} >0$) with the integer $n\approx
|U_{BF}|/U_{BB}$ can form. In recent experiments \cite{ETH1,Ospelkaus2006}
complexes of the type FB$_n$ with $n$ as large as $n=3$ may have played an
important role.

Depending on the parameters, the effective interaction between composite
fermions can be tuned to become attractive or repulsive. In the case of
effective attractive interaction, a p-wave superfluid of pairs of composite
fermions will form \cite{SCF,Efremov2002}. For effective repulsive
interactions and low filling of composite fermions the ground state is a
Fermi liquid while near half filling insulating states may appear
\cite{Lewenstein2004}. For clarity, we will limit ourselves here to the case
of composite fermions forming a weakly interacting Fermi liquid with a well
defined Fermi surface.

A pure quasi-molecular phase FB$_n$ will only exist when the number of
bosons N$_B$ and fermions N$_F$ is commensurate, i.e. $N_B=nN_F$.
Conversely, in the case of a pure phase of composites of the type
F$\overline{\rm B}_n$ the number of bosons must be $N_B=N\cdot k-N_F\cdot
n$, with the integer $k \geq n$ \cite{note2} and $N$ standing for the total
number of the optical lattice sites. Free fermions or bosons will be present when
these conditions are not fulfilled, or when going away from the strongly
interacting regime, where the composites can partly dissociate. Then two
types of Fermi surfaces can coexist --- one for the composites 
 and the other for the original fermions. Moreover, a
portion of the B-bosons can form a BEC \cite{Sachdev}.

Here we will consider the limit of low density
of the composites, so that
the size of one molecule $a_m$ is much less than the average
distance $r_o$ between them. In this case the description becomes quite straightforward in terms
of the centers of mass of the composites only.
As $a_m/r_o$ increases and eventually becomes comparable to 1, the composite fermions will
 dissociate according to the scenario \cite{Sachdev}.

\subsection{Detection in weakly interacting Bose-Fermi mixtures}

In the weakly interacting Bose-Fermi mixture without composites present,
practically all important information can be deduced by imaging the bosonic
$n_B({\bf q})$ and fermionic $n_F({\bf q})$ quasimomentum distributions. In
this method, the quasimomentum distribution is first mapped to the momentum
distribution by removing the lattice potential on a time scale which is slow
with respect to single particle physics but fast compared to quasimomentum
changing collisions. After subsequent free expansion for a time $t$ the
quasimomentum distribution is detected by imaging the density
\cite{Greiner2001,Koehl2004}. Apart from unimportant multiplicative factors,
$n_B({\bf q})\approx \int d{\bf x} \exp(-i{\bf q}{\bf x})\langle
\psi^\dagger_B({\bf x})\psi_B(0)\rangle$ and $n_F({\bf q})\approx \int d{\bf
x} \exp(-i{\bf q}{\bf x})\langle \psi^\dagger_F({\bf x})\psi_F(0)\rangle$,
where $\psi_B({\bf x})$ and $\psi_F({\bf x})$ are the bosonic and fermionic
field operators, respectively, just before the lattice was removed and the
averaging $\langle ... \rangle$ is taken over the corresponding state in the
lattice. In the weakly interacting regime, the bosonic
 quasi-momentum distribution reflects the off-diagonal long range
order (ODLRO) inherent in the one-particle density matrix $\rho_B({\bf
x},{\bf x}')=\langle \psi^\dagger_B({\bf x})\psi_B({\bf x}')\rangle$,
resulting in a very small spread of quasi-momenta. The fermionic
one-particle density matrix $\rho_F({\bf x},{\bf x}')=\langle
\psi^\dagger_F({\bf x})\psi_F({\bf x}')\rangle$ features Friedel
oscillations which are due to the sharp edge in the Fermi distribution $n_F({\bf
q})$ (at $T=0$): $n_F({\bf q}) \approx 1$ for $|{\bf q}| < q_F$
and $n_F({\bf q}) \approx 0$ for $|{\bf q}| > q_F$, where $q_F$ stands for
the Fermi wavevector.

\subsection{Direct detection of composites}\label{com}

The method described above is not applicable to composite fermions, since
they typically dissociate during the expansion. However, the center of mass
momentum ${\bf Q}$ of a FB$_n$ atomic complex is conserved during time of
flight. Therefore the momenta of all its n+1 constituents after expansion
are directly correlated and add up to ${\bf Q}$. We point out that the
observation of such correlations constitutes direct evidence for the
existence of the composites. In the following we will describe how these
correlations can be used to directly detect the (quasi)momentum distribution
of the composites in the lattice.

For an ultracold gas that contains many composites, the fermionic (bosonic)
momentum distributions, i.e. the absorption images after time of flight,
represent the momenta of many fermions (bosons). Consequently, the
correlations between the constituents originating from one composite are not
easily visible, since it is not clear which of the many bosons and fermions
made up one composite. The correlations can therefore not be determined from
the form of the momentum distributions but only from the atomic shot noise
visible in them. Since the correlations are between one fermion and $n$
bosons, the corresponding correlator is $\langle \delta n_F({\bf X}_1)\delta
n_B({\bf X}_2)...\delta n_B({\bf X}_{n+1})\rangle$. To determine the amount
of correlation, this correlator must be integrated over $n+1$ coordinates
under the constraint that the center of mass momentum {\bf Q} of the
constituents is fixed:
\begin{eqnarray}
I_{n+1}({\bf Q})= \int d{\bf q}_2...d{\bf q}_{n+1} \quad \quad
\label{I_n+1} \\
\langle : \delta n_F({\bf q}_1)\delta n_B({\bf q}_2)...\delta n_B({\bf
q}_{n+1}):\rangle, \nonumber
\end{eqnarray}
with ${\bf Q}= {\bf q}_1 + {\bf q}_2+...+ {\bf q}_{n+1}$ and ${\bf
q}_1=m_F{\bf X}_1/t,\, {\bf q}_2=m_B{\bf X}_2/t,..., {\bf q}_{n+1}=m_B{\bf
X}_{n+1}/t$. Here $: ... :$ implies the normal order of the operators ---
when all the creation operators stay to the left of the annihilation ones.
The averaging $\langle ... \rangle$ is performed over a particular realization
of a many body eigenstate $|\Psi \rangle$ and, then, over many realizations.
In appendices A and B we show that the amount of correlation
$I_{n+1}({\bf Q})$ is proportional to the number of composites with given quasi-momentum $\bf Q$, that is
\begin{equation}
I_{n+1}({\bf Q})\sim \tilde{\rho}_{CF}({\bf
Q}) - C,\label{I_equals_cc}
\end{equation}
where $\tilde{\rho}_{CF}({\bf
Q})=\langle c^{\dagger}({\bf Q}) c({\bf Q})\rangle$ and $c^{\dagger}({\bf Q}),c({\bf Q})$ are the composite fermion creation
and annihilation operators in the momentum space and $C$ is a
constant term. To summarize, the quasimomentum distribution $\tilde{\rho}_{CF}({\bf
Q})$  of the composites can be found by
measuring the correlations between experimental absorption images
according to the equation (\ref{I_n+1}).

The relation (\ref{I_equals_cc}) is proved rigorously in appendices A and B.
It is instructive to follow the main steps of the proof: The operators
\begin{eqnarray}
\delta n_F(\textbf{X})&=&\psi^\dagger_F(\textbf{X})\psi_F(\textbf{X}) - \langle
\Psi|\psi^\dagger_F(\textbf{X})\psi_F(\textbf{X})|\Psi\rangle,
\label{DF} \\
\delta n_B(\textbf{X})&=&\psi^\dagger_B(\textbf{X})\psi_B(\textbf{X}) - \langle \Psi |
\psi^\dagger_B(\textbf{X})\psi_B(\textbf{X})|\Psi\rangle, \label{DB}
\end{eqnarray}
appearing in equation (\ref{I_n+1}) are expressed as functions of the onsite
fermionic and bosonic creation and annihilation operators
$f^{\dagger}_i,f_i,b^{\dagger}_i,b_i$, with $i$ being the site
index, in the lattice before expansion with
the help of eqs.(\ref{op_FF},\ref{op_BB},\ref{WWW}). Inserting these
expressions into eq.(\ref{I_n+1}) yields sums over terms with $2(n+1)$
creation and annihilation operators on different sites. Many of these can be
set to zero, resulting in
\begin{eqnarray}
I_{n+1}({\bf Q})&\sim& \sum_{ij}{\rm e}^{i{\bf Q}({\bf x}_i-{\bf x}_j)} \langle [
c^\dagger_i c_j -\delta_{ij}c^\dagger_i c_i]\rangle,
 \label{I_2_op}
\end{eqnarray}
where $c^\dagger_i,\, c_i$ are the onsite composite fermion creation-annihilation
operators as defined in eq.(\ref{compo}) of the Appendix \ref{phys}, and ${\bf x}_i$ denotes
the site $i$ coordinates.
The first term in brackets in eq.(\ref{I_2_op}) is $\tilde{\rho}_{CF}({\bf
Q})$: as can be noticed, after introducing the composite Fermi operators in the Fourier space $c({\bf
Q})\sim \sum_i \exp(-i{\bf Q} {\bf x}_i) c_i$, one finds equation
(\ref{I_equals_cc}). While $\tilde{\rho}_{CF}({\bf
Q})$ gives
the occupation number of composites with quasimomentum \textbf{Q}, 
the second term is the constant $C$   which takes care of the normalization $
\int d{\bf Q}\,I_{n+1}({\bf Q})=0.$ The last relation follows from the
observation that the integral $\int d{\bf Q}\,I_{n+1}({\bf Q})$ is exactly
$\langle  \delta N_F  :(\delta N_B)^n:  \rangle$ where $\delta N_{B,F}$ denote
fluctuations of the total numbers of bosons and fermions $N_{B,F}$, respectively.
Since $N_F$ is
diagonal on any many-body eigenstate $|\Psi \rangle$, such a mean is exactly
zero for any particular realization of $|\Psi \rangle$, that is $\delta N_F |\Psi \rangle=0$. Both contributions
in eqs.(\ref{I_equals_cc},\ref{I_2_op})
can easily be distinguished because they have very different structures: the
first one is concentrated within the Fermi surface of the composite fermions
and the second is spread uniformly over all momenta in the first Brillouin
zone.

In the case of the complexes F$\overline{\rm B}_n$ ( see the discussion in Appendix \ref{strong}
below eqs.(\ref{space_BF_h},\ref{space_BF_hc})) the only
change is in the definition of the momentum $\bf Q$ in eqs.(\ref{I_n+1}). It
becomes ${\bf Q}= {\bf q}_1 - {\bf q}_2-...- {\bf q}_{n+1}$.

As discussed in ref.\cite{Sachdev}, decreasing the F-B interaction $U_{BF}$
 will eventually result in coexistence of {\it three} types of quasi-particles: some portion
of free B-bosons and F-fermions and the composites.
In order to detect the composites in this case, the noise must be
measured on  top of signals from the structured backgrounds $\langle
n_F({\bf q})\rangle, \,\, \langle n_B({\bf q})\rangle $ in the fermionic and
bosonic channels.

\subsection{Correlated versus uncorrelated noise}\label{noise}
Here we discuss the detection requirements under the assumption that the
atomic quantum and shot-to-shot noise is dominant. The above results
indicate that measuring the noise in the density correlator of the $n+1$
order can provide a one-particle density matrix for the centers of mass of
the composites. It is important, however, that the uncorrelated noise due to
photon shot noise, inaccuracy $\Delta $ in determination of the number of
particles in each detection bin, etc., is much less than the correlated
noise in one shot. Otherwise, many shots $S$ are required in order to reduce
the effect of the uncorrelated noise (by the factor $\sim 1/\sqrt{S}$). Here
we will show that, while for diatomic composites the uncorrelated noise
(with respect to the correlated one) is insignificant for large enough
numbers $N_F\sim N_B$ (decays as $1/\sqrt{N_{F}}$), for multi-atomic
composites FB$_n$ with $n=3,..$ it grows as   $\sim N_{F}^{n/2 -1}$, with
the case $n=2$ being marginal. This makes the direct scheme of measuring the
density matrix of the $n>2$ composites impractical because the required
number of shots grows as $S \sim N_{F}^{n-2}$. This effect is caused by a
strong increase of the phase space with $n$ when the composites are
dissociated as a result of the lattice release.

In what follows we will normalize the correlated and uncorrelated noises by
the corresponding measured signal
\begin{eqnarray}
 I_{sig}=&\int&d{\bf X}_2...d{\bf X}_{n+1}\langle n_F(\bf{Q} - \bf{q}_2-...-\bf{q}_{n+1})\rangle
\nonumber \\
 &\,&\langle n_B(\textbf{q}_2)\rangle...\langle n_B(\textbf{q}_{n+1})\rangle
\label{signal_n}
\end{eqnarray}
where ${\bf q}_k=m_B {\bf X}_k/t$. This integral can be estimated as $I_{
sig} \approx N_B^n N_F/N_{\text{bin}}=n^n N_F^{n+1}/N_{\text{bin}}$, where
we consider exact matching $N_B=nN_F$ and $N_{\text{bin}}$ stands for the
total number of the detection bins. In the following estimates we will be using
the relations $N_B\sim N_F \sim N$, so that the actual numbers of particles 
can be replaced by the total number of sites (provided the densities are finite).

The uncorrelated noise is caused by technical inaccuracies $\Delta ({\bf
X}_i)$ in the detection of the number of, e.g., fermions in each detection
bin.  The effect of such noise becomes critically magnified for large $n$
even under the condition that $|\Delta ({\bf X}_i)|$ is much smaller than a
typical quantum fluctuation  $\sim \sqrt{N_F/N_{\text{bin}}} \sim \sqrt{N/N_{\text{bin}}}\gg 1$ of the number
of detected particles per each bin. We note that, under the ideal condition
$\Delta ({\bf X}_i)=0$, quantum fluctuations of the particles number in each
bin do not give rise to any uncorrelated noise in the correlator
(\ref{I_n+1}). This is due to the fact that the center of mass momenta of
the composites are good quantum numbers -- in contrast to the momenta of the
constituents -- and therefore the corresponding quantity, the density of the
composites represented by the operator in eq.(\ref{I_n+1}) does not
fluctuate in any particular many-body eigenstate $|\Psi \rangle$.  If,
however, the measurement is done with an error $\Delta ({\bf X}_i)$, the
operator determining the correlator (\ref{I_n+1}) acquires the contribution
$ \Delta \hat{I}= \int d{\bf X}_2...d{\bf X}_{n+1} \Delta({\bf X}_1)\delta
n_B({\bf X}_2)...\delta n_B({\bf X}_{n+1})$, which can be considered as a
fair estimate of the total error under the condition $|\Delta| \ll
\sqrt{N_F/N_{\text{bin}}}$. Clearly the mean of it is zero, $ \langle \Delta
\hat{I}\rangle=0$. However, its {\it rms} fluctuation $E=\sqrt{\langle
\Delta \hat{I} \Delta \hat{I}\rangle}/I_{sig}$ (normalized by $I_{sig}$) is
not zero. We estimate: $\langle \Delta \hat{I} \Delta \hat{I}\rangle \sim
\langle \Delta^2\rangle \left(\int d{\bf X}_1 \int d{\bf X}_2 \langle \delta
n_B({\bf X}_1)n_B({\bf X}_2)\rangle \right)^n \sim \langle \Delta^2\rangle
N^n$, where we have taken into account that the quantum fluctuation of the
number of bosons in each bin is $\sim \sqrt{N/N_{bin}}$ and we have
considered no correlations between the bins (and omitted factors dependent on $n$ like
$n!$, $n^n$, etc.) . Given this estimate and the one
for $I_{sig}$ in eq.(\ref{signal_n}), we find
\begin{equation}
E= \frac{\langle|\Delta|\rangle N_{\text{bin}}}{N^{1+n/2}}, \label{E}
\end{equation}
where $\langle|\Delta|\rangle$ stands for the {\it rms}-fluctuations of  $\Delta ({\bf X}_i)$ .

The quantity $E$ charaterizes the uncorrelated noise contribution in one shot. If $S$
shots are performed, it is reduced as $E \to E/\sqrt{S}$ so that, in
principle, the effect of the uncorrelated noise can be reduced below the
correlated contribution $I_{n+1}/I_{sig}$ after taking a sufficient amount
of the shots. In reality, the required $S$ must scale not faster than $
S\sim {\cal O}(1)$ with $N_{F}$.
 The correlated contribution (\ref{I_2_op})
is given just by the total number of the composites $\sim N_F$ in each bin:
$I_{n+1} \sim N_F/N_{bin}  \sim N/N_{bin}$. Hence,
 $I_{n+1}/I_{sig} \approx N^{-n}$ and $\Delta I/I_{n+1} \sim \langle |\Delta |\rangle N_{bin} N^{n/2-1}/\sqrt{S}$.
Thus,
the required number of shots is
\begin{equation}
S \gg |\Delta|^2 N_{\text{bin}}^2 N^{n-2}. \label{S}
\end{equation}
For $n=1$, this condition can be achieved for large enough $N$ and not
very large number of bins $N_{\text{bin}}$. For $n=2$, the condition is
marginally reachable, and, for $n>2$, it becomes impossible to fulfill.

It is worth mentioning that, in the case of composite bosons, say, AB$_n$,
where A labels some boson of a  kind different from B, the condition
(\ref{S}) becomes less strict. Indeed, $I_{n+1}$ (\ref{I_n+1}) in this case
acquires an additional factor $\sim N$ due to the ratio of the
correlation volumes of a boson and a fermion.
Indeed, the sum $\sum_{ij}{\rm e}^{i{\bf Q}({\bf x}_i - {\bf x}_j)} \langle
c^\dagger_i c_j\rangle$ can be estimated as $\sim N \xi^3$, where
$\xi $ is a typical distance on which the correlator $\langle
c^\dagger_i c_j\rangle$ becomes zero. For fermions it is given
by the interparticle separation $r_o$. For bosons in BEC, it is the size of the
lattice and one finds $I_{n+1} \sim N^2/N_{bin}$.
Accordingly, keeping the signal and the uncorrelated noise estimates the
same, we find the number of the required shots
$S
\gg |\Delta|^2 N_{\text{bin}}^2 N^{n-4}$. This implies that
 for composite bosons the method described above can
be used if $n \leq 4$, that is the total number $n+1$ of the constituents
should not exceed five.

Concluding this section, we mention that performing shot-to-shot averaging
should be done with care, because, otherwise, the noise scaling with $N$
can become even stronger. The operators of fluctuations $\delta n_F = n_F -
\langle n_F\rangle,\, \delta n_B = n_B - \langle n_B\rangle,\, \delta n_A =
n_A - \langle n_A\rangle$ are defined with respect to averaging over a
particular ground state characterized by given total numbers of atoms $N_F, N_B, N_A$. In other
words, the average number of atoms $ \langle n_F\rangle,\, \langle
n_B\rangle,\,\langle n_A\rangle$ per each bin must be determined in each
shot. Then, the final value $I_{n+1}$ can be averaged over the shots.
Otherwise, shot-to-shot fluctuations of $N_{F,B,A}$ which typically scale at
least as $\sqrt{N_{F,B,A}}$, respectively, will introduce stronger noise if
the means $ \langle n_F\rangle,\, \langle n_B\rangle,\,\langle n_A\rangle$
are understood as the total ensemble means. In the latter case the difference between
the mean in one particular realization and the total one serves as an
additional uncorrelated noise.
Then, one can estimate $\Delta \sim \sqrt{N}/N_{bin}$ and, accordingly,
find  $\Delta I/I_{n+1} \sim  N^{(n-1)/2}/\sqrt{S}$, so that
the required number of shots scales as $S \sim N^{n-1}$, implying that
the detection of the composite fermions with $n>1$ is not possible.
Similarly, for the composite bosons: $\Delta I/I_{n+1} \sim  N^{(n-3)/2}/\sqrt{S}$,
and the detection for $n>3$ becomes impossible.

\section{Indirect indication of the composite fermions in the bosonic density-density noise
correlator}\label{indirect}

Here we discuss the situation when all bosons are bound to be
the constituents of the composite fermions so that there is
no ODLRO in the lattice. The question is how the
bosonic density-density correlator
\begin{eqnarray}
I_{bb}(\textbf{Y})&=&\int d{\bf X}\langle \Psi|:n_B(\textbf{X}+{\bf Y}/2)
n_B(\textbf{X}-{\bf Y}/2):|\Psi \rangle,
\nonumber \\
\label{bb}
\end{eqnarray}
(taken in the normal order) is affected by such state. Below we show that
(\ref{bb}) exhibits the Bragg structure similar to that observed in the Bose
Mott insulator (MI) \cite{Bloch_noise,Spielman2006}. The relations
(\ref{fff},\ref{bbb}) in the physical space of the composites (\ref{C})
allow expressing the correlator (\ref{bb})
 as the mean over the operator
\begin{eqnarray}
\hat{I}_{bb}(\textbf{Y})&=&n^2\sum_{ij}(1+{\rm e}^{i(m_B{\bf Y}/t)({\bf x}_i-{\bf x}_j)}) c^\dagger_ic^\dagger_jc_jc_i
\nonumber \\
&+&n(n-1)\sum_i c^\dagger_i c_i,
\label{BB}
\end{eqnarray}
where the last term, which is $\sim  o(1/N)$ with respect to the first one,
can be ignored. 
Taking the means and replacing $\langle
c^\dagger_ic^\dagger_jc_jc_i\rangle=\nu^2(1-\delta_{ij})$, with $\nu $ being
the average onsite population of fermions, we find that eq.(\ref{BB}) has
exactly the same structure observed for the bosonic Mott insulator
\cite{Bloch_noise,Spielman2006}: the  term $\sim \exp [i(m_{F,B}{\bf
Y}/t)({\bf x}_i-{\bf x}_j)]$ is the correlated contribution \cite{HBT} which
peaks as long as ${\bf Y}$ matches the positions of the Bragg peaks (in
other words, ${\bf Y}=0$ {\it modulo} the primitive vectors of the reciprocal lattice).
 It is
important that this structure exists for {\it any} filling factor of the
composites $\nu \leq 1$, that is, it is not a sole property of the Mott
insulating ground states.

Regarding the fermionic correlator
\begin{eqnarray}
I_{ff}(\textbf{Y})=\int d{\bf X}\langle \Psi|:n_F(\textbf{X}+{\bf Y}/2)
n_F(\textbf{X}-{\bf Y}/2):|\Psi \rangle,
\nonumber \\
\label{ff}
\end{eqnarray}
we note that it must exhibit the antibunching peaks, that is,
$I_{ff}(\textbf{Y})=0$ for $\textbf{Y}=0$ ({\it modulo} the primitive vectors of the reciprocal lattice)
regardless of the nature of the ground state because the normal product
$:n_F(\textbf{X}) n_F(\textbf{X}):$ is zero (due to the Pauli principle and
the periodicity of the lattice). This can be verified directly with the help
of eqs.(\ref{op_FF},\ref{WWW}). Then, Eq.(\ref{ff}) becomes a periodic
function $I_{ff}(\textbf{Y})$, with the periods determined by the primitive vectors of the reciprocal lattice, and one finds
   $I_{ff}(0)\sim \int d{\bf q} \sum_{ijkl}\exp(i{\bf q}({\bf x}_i + {\bf x}_k - {\bf x}_j - {\bf x}_l))
\langle f^\dagger_i f^\dagger_k f_lf_j \rangle =0 $ due to the Fermi statistics: $f_lf_j + f_jf_l=0$.

The physical reason why the bosonic correlator (\ref{bb}) shows bunching
peaks (\ref{BB}) in a strongly interacting mixture stems from a very general
principle, which is best explained by starting from the opposite case. In
any non-interacting {\it pure} quantum mechanical state the correlator
$\langle \Psi_0| \delta n_{B}(\textbf{X}) \delta n_{B}({\bf Y})|\Psi_0 \rangle$
(not taken in the normal order!) must be exactly zero. To understand this
one must remember that the density operators in the far zone are equivalent
to the (quasi-)momentum operators in the lattice and that non-interacting
systems are characterized by well defined quasi-momenta. Thus, there are no
fluctuations in each particular many-body (non-interacting) eigenstate
$|\Psi_0\rangle$ and the correlator $\langle \Psi_0| \delta
n_{B}(\textbf{X}) \delta n_{B}({\bf Y})|\Psi_0 \rangle$ is zero. In other
words the operator (\ref{DB}) is {\it zero} on space of states $|\Psi_0
\rangle$ of the ideal systems: that is, $\delta n_{B}(\textbf{X})|\Psi_0
\rangle=0$. In the thermodynamical limit at low $T$ the normal ordering
in eq.(\ref{bb}) can be ignored for bosons (as opposed to fermions!).
Consequently, a weakly interacting BEC of B-atoms will lead to essentially uniform
distributions in eq.(\ref{bb}) \cite{note_T}.

Meanwhile in a quasi-molecular phase of strongly interacting Bose-Fermi
mixtures the momenta of the original bosons are undefined in each {\it pure}
many-body state $|\Psi \rangle$ in complete analogy with the MI.
 Accordingly, the densities in the far zone fluctuate strongly and
produce the bunching behaviour in (\ref{BB}). In this case quantum
fluctuations are responsible for the Hanbury Brown and Twiss effect. Setting
it more technically, in each many-body eigenstate the correlator $\langle
\Psi| \delta n_{B}(\textbf{X}) \delta n_{B}({\bf Y})|\Psi \rangle$ is
proportional to the interaction vertex \cite{Landau} and, if the interaction
is small, the correlator is, practically, zero. In general, having a strong
interaction regime is the requirement for observing the distinct patterns
(\ref{BB}) in pure states. This condition is more general than the one for
having energy gap with respect to single-particle excitations
\cite{Bloch_noise}. In the state mentioned in Sec.\ref{com} where all
three types of quasi-particles are present, such a gap is zero and the
composites can still be well defined. Accordingly, the Hanbury Brown and
Twiss pattern (\ref{BB}) should be observable. Obviously, some care should
be taken in order not to confuse the pattern due to the composites with that
caused by thermal fluctuations \cite{note_T}.

\section{Conclusion and acknowledgement}
We have suggested the extended density-density correlator method for the
direct imaging of the composites in optical lattice and have evaluated its
feasibility for multi-atomic fermionic and bosonic complexes. The method is
expected to work well for fermionic composites consisting of less than four
atoms and for bosonic ones consisting of less than six atoms. Above these
numbers, the uncorrelated noise becomes too large.

As an indirect method for detecting the composites, measuring the bosonic
density-density correlators must reveal the typical Hanbury Brown and Twiss
structure first observed for the bosonic Mott insulator. We point out that
this observation is a generic consequence of strong interactions in a
two-component mixture in optical lattices.

We thank M. K\"ohl for a critical reading of the manuscript. This work was
supported by NSF grant PHY-0426814, PSC-CUNY grant PSC-66556-0037.

\appendix
\section{Strongly bound FB and F$\overline{\rm \textbf{B}}$ pairs}\label{strong}

Here we will give a more extended treatment of the FB case discussed in the
introduction. It is important that the correlator (\ref{I_2}) turns out to
be proportional to the Fourier representation of the {\it in-situ} two-body
density matrix $\langle \psi_F^\dagger({\bf x}_1)\psi_B^\dagger({\bf
x}_2)\psi_F({\bf y}_1)\psi_B({\bf y}_2)\rangle$ with respect to the
coordinate ${\bf x}_1 - {\bf y}_1$. As a matter of fact, since a
boson can only be found close to a fermion, one can set ${\bf x}_1={\bf
x}_2,\,{\bf y}_1={\bf y}_2$ in this correlator without affecting
its low energy properties (as long as $a_m/r_o \ll 1$).

In the following we will justify the statement
$I_{CF}(\textbf{Q})\sim\tilde{\rho}_{CF}(\textbf{Q})$. We assume that the
wavefunction of relative motion $\Phi ({\bf r})$ of a boson and a fermion in
a pair is of the s-wave type $\Phi ({\bf r})\approx \exp( -|{\bf
r}|/a_m)/a_m^{3/2}$ with $a_m$ being of the order of the lattice constant.
[In the case of the non-s-wave BF pairing, the correlator (\ref{I_2})
vanishes after integration over, say, ${\bf q}_2$ with $\bf Q$ kept
constant. We do not consider such an exotic possibility]. We will see in the
following that once $|{\bf Q}\,a_m| \ll 1$, the extent of the wavefunction
$a_m$ has no effect on $I_{CF}(\textbf{Q})$ or
$\tilde{\rho}_{CF}(\textbf{Q})$.
In order to calculate $I_{CF}(\textbf{Q})$, we expand the fermionic and
bosonic operators in the optical lattice as
\begin{eqnarray}
\psi_F({\bf x})&=&\sum_i W_0({\bf x}-{\bf x}_i)\,f_i,
\label{op_F} \\
\psi_B({\bf x})&=&\sum_i W_0({\bf x}-{\bf x}_i)\,b_i,
\label{op_B}
\end{eqnarray}
where
$W_0({\bf x}-{\bf x}_i)$ stands for the Wannier function
located at $i$th site. For simplicity, we consider $W_0({\bf x})$ to be the same for fermions
and bosons. Upon free expansion of a particle of
mass $m$, the Wannier function in the far zone becomes (apart from a
numerical coefficient)
\begin{eqnarray}
W_t({\bf X}-{\bf x}_i)= \left(\frac{m}{t}\right)^{3/2}{\rm e}^{-i{\bf
q}{\bf x}_i+ im{\bf X}^2/2t}\,\,\tilde{W}({\bf q}) \nonumber\\
\label{WWW}
\end{eqnarray}
 where
${\bf q}=m{\bf X}/t$ (we employ units in which $\hbar =1$) and
$\tilde{W}({\bf q})=\int d{\bf x}' \exp(-i{\bf q}{\bf x}') W_0({\bf x}')$ stands for the Fourier transform of the Wannier
function. In the far zone the operators (\ref{op_F},\ref{op_B}) become
\begin{eqnarray}
\psi_F({\bf X})&=&\sum_i W_t({\bf X}-{\bf x}_i)\,f_i,
\label{op_FF} \\
\psi_B({\bf X})&=&\sum_i W_t({\bf X}-{\bf x}_i)\,b_i,
\label{op_BB}
\end{eqnarray}
where the corresponding mass $m_F, \, m_B$ is replacing $m$.

Then, substitution of eqs.(\ref{op_FF},\ref{op_BB})
into the definition of the densities in the far zone (\ref{DF},\ref{DB}) and
employing them in eq.(\ref{I_2})
give
\begin{widetext}
\begin{eqnarray}
I_{CF}({\bf Q})&=&\int d{\bf q}_2 \frac{m^3_F|\tilde{W}({\bf Q}-{\bf q}_2)\tilde{W}({\bf q}_2)|^2}{t^3} 
[\langle f^\dagger({\bf Q}-{\bf q}_2)b^\dagger({\bf q}_2)f({\bf Q}-{\bf q}_2)b({\bf q}_2)\rangle -
\nonumber \\
&\,& \langle f^\dagger({\bf Q}-{\bf q}_2)f({\bf Q}-{\bf q}_2)\rangle \langle b^\dagger({\bf q}_2)b({\bf q}_2)\rangle]
 \label{I_2_full} 
\end{eqnarray}%
\end{widetext}
where 
$f({\bf q})=\frac{1}{\sqrt{N}}\sum_i \exp(-i{\bf q}{\bf x}_i)f_i$ and
$b({\bf q})=\frac{1}{\sqrt{N}}\sum_i \exp(-i{\bf q}{\bf x}_i)b_i$ are the Fourier representations
of the onsite operators.
We note that
$I_{CF}({\bf Q})$ scales with the expansion radius $R\sim t$
as ~$ \sim 1/R^{3}$, that is, as though it is just a density
of some particle undergoing free expansion.

The momenta involved in $\tilde{W}({\bf q})$ are of the order of the inverse
lattice constant $a_l$. In fact, as the lattice is being ramped down
adiabatically with respect to the single particle states,
 the Wannier function undergoes a transformation from
that corresponding to the deep lattice
to the one in the very shallow lattice. In the latter case, $W({\bf x})$ can
be found explicitly from the definition (1) of ref.\cite{Wannier}, where for
the Bloch function one can use just the exponential $\exp(i{\bf k}{\bf x})$.
Then, the square of the Fourier transform $|\tilde{W}({\bf q})|^2$ trivially
becomes the volume of the elementary cell $\Omega_{BZ} \approx a_l^3$ when
$\bf q$ belongs to the first Brillouin zone and zero otherwise. Accordingly,
the integration $\int d{\bf q}_2...$ in eq.(\ref{I_2_full}) proceeds over
the first Brillouin zone $\int_{BZ} d{\bf q}_2...$ with the factor
$|\tilde{W}({\bf Q}-{\bf q}_2)\tilde{W}({\bf q}_2)|^2=1$ as long as the
typical values of $\bf Q$ are much smaller than the inverse lattice constant
$a_l$ (for ${\bf Q}$ being outside the first zone, this factor is
essentially zero, if the Fermi "sphere" has a volume $\approx Q^3_{CF}$ much
less than $1/\Omega_{BZ}$).
In particular, this  implies that the "imaged" composites are
localized mostly within the first Brillouin zone similarly to the rampdown
procedures \cite{Greiner2001,Koehl2004} used for the atomic imaging.  As can
be seen, the spreading over the next zone is suppressed by the factor
$Q^3_{CF}\Omega_{BZ} \ll 1$ due to the small fraction of the phase volume in
the integral (\ref{I_2_full}) which allows access to the center of the next
zone. 

If bosons and fermions do not bind together, the correlator
in eq.(\ref{I_2_full}) is essentially zero. If, however, there are tightly bound
composites FB in the lattice, the correlator  
become finite whenever a boson is found within the proximity $a_m$ of a fermion.
This is indicated in a choice of the physical space 
\begin{eqnarray}
&\,&|1,\{n_{\bf Q}\}\rangle =\prod_{\bf Q} (c^\dagger_{\bf Q})^{n_{\bf Q}}|0\rangle, 
\label{space_BF} \\
c^\dagger_{\bf Q}&=&\int d{\bf q} \, \tilde{\Phi}({\bf q})f^\dagger({\bf Q}/2 +{\bf q})b^\dagger ({\bf Q}/2 -{\bf q}),
\label{space_BFC}
\end{eqnarray}
with $n_{\bf Q}=0,1$, of weakly interacting composite fermions (in the considered limit $r_o \ll a_m$).
It is represented by the composite creation-annihilation operators
$c^\dagger_{\bf Q}, c_{\bf Q}$ of a BF pair with total momentum $\bf Q$ in terms
of the Fourier components of the fermionic and bosonic operators and of the relative
wavefunction $\tilde{\Phi}({\bf q})$. It is important that
the low energy properties do not depend on details of $\tilde{\Phi}({\bf q})$ in the limit
of low density of the composites. This can be seen
by direct calculation of eq.(\ref{I_2_full}) over the state (\ref{space_BF}).
Specifically, the state $f({\bf Q}-{\bf q}_2)b({\bf q}_2)|1,\{n_{\bf Q}\}\rangle$
contains contribution $\tilde{\Phi}({\bf Q}/2- {\bf q}_2)|0\rangle$ coming from the action of 
$f({\bf Q}-{\bf q}_2)b({\bf q}_2)$ on the term
describing just single composite $c^\dagger_{\bf Q}|0\rangle$ without affecting the others. It also
contains the exchange terms involving pairs of the composites with different momenta. These
exchange terms, however, can clearly be neglected because they contribute factors $ (a_m/r_o)^3 \ll 1$.   
Thus, neglecting these terms, we find
$\int_{BZ} d{\bf q}_2\,\,\langle \{n_{\bf Q}\}, 1| f^\dagger({\bf Q}-{\bf q}_2)b^\dagger({\bf q}_2)f({\bf Q}-{\bf q}_2)b({\bf q}_2)|1,\{n_{\bf Q}\}\rangle=
\int_{BZ} d{\bf q}_2 |\tilde{\Phi}({\bf q}_2)|^2\,\, n_{\bf Q}=\langle \{n_{\bf Q}\}, 1| c^\dagger({\bf Q})c({\bf Q})|1,\{n_{\bf Q}\}\rangle$, since $\int_{BZ} d{\bf q}_2 |\tilde{\Phi}({\bf q}_2)|^2=1$
as the normalization condition. Thus, in the physical space (\ref{space_BF},\ref{space_BFC})
the operator $\int_{BZ} d{\bf q}_2 f^\dagger({\bf Q}-{\bf q}_2)b^\dagger({\bf q}_2)f({\bf Q}-{\bf q}_2)b({\bf q}_2)$
is equal to $c^\dagger({\bf Q})c({\bf Q})$ and its averaging produces the momentum distribution
$\tilde{\rho}_{CF}({\bf Q})= \langle \{n_{\bf Q}\}, 1| c^\dagger({\bf Q})c({\bf Q})|1,\{n_{\bf Q}\}\rangle$
of the composite fermions as discussed in the Introduction below eq.(\ref{I_2}).

As can be directly checked, the uncorrelated contribution
$\int_{BZ} d{\bf q}_2\langle f^\dagger({\bf Q}-{\bf q}_2)f({\bf Q}-{\bf q}_2)\rangle \langle b^\dagger({\bf q}_2)b({\bf q}_2)\rangle$ in eq.(\ref{I_2_full}) is independent of $\bf Q$ as long as $|{\bf Q}| \ll a^{-1}_m$ and, thus,
can be replaced by a constant $C$. This constant can be restored from the normalization on the density
of fermions (and bosons).  Finally, the expression (\ref{I_2}) becomes 
$I_{CF}({\bf Q}) \sim \tilde{\rho}_{CF}({\bf Q}) - C$, which is a particular case of a more
general eq.(\ref{I_equals_cc}) discussed in the section \ref{com}.

We also comment on the case of purely repulsive interactions when the
pairing occurs between fermions and bosonic holes \cite{SCF}. 
The corresponding physical space, then, becomes
\begin{eqnarray}
&\,& |1,\{n_{\bf Q}\}\rangle =\prod_{\bf Q} (c^\dagger_{\bf Q})^{n_{\bf Q}}\prod_i (b^\dagger_i)^k|0\rangle 
\label{space_BF_h} \\
c^\dagger_{\bf Q}&=&\int d{\bf q} \, \tilde{\Phi}({\bf q})f^\dagger({\bf Q}/2 +{\bf q}) b({\bf q}-{\bf Q}/2),
\label{space_BF_hc}
\end{eqnarray}
where $k=1,2,..$ and $\tilde{\Phi} ({\bf q}) $ is now a wavefunction of relative motion of a
fermion and bosonic hole. 
Accordingly, the total momentum $\bf Q$ is now the {\it difference}
of the momenta carried by the fermion ${\bf Q}/2 +{\bf q}$ and by the bosonic
hole ${\bf q}-{\bf Q}/2$.
Then, as discussed in the Introduction, the momentum $\bf Q$ in eq.(\ref{I_2})
is to be replaced by the {\it difference}
of the momenta: ${\bf Q}_-={\bf q}_1 - {\bf q}_2$.

\section{The physical Hilbert space for strongly bound complexes FB$_n$}\label{phys}
Insensitivity of the low energy physics to the internal structure of the
composites discussed in the Appendix \ref{strong} for the case FB (as an example)
 allows a simplification
of the analysis by projecting the Hamiltonian as well as any measurable
quantity into states having either no particles on a site or $1+n$ particles
--- one fermion and $n$ bosons. In other words, the constituents of a
composite are located at the same site and the probability to find an
unbound boson or fermion on another site is vanishingly small. This means
that the physical operators are
\begin{equation}
c_i=\frac{f_ib^n_i}{\sqrt{n!}},\quad c^\dagger_i=\frac{f^\dagger_ib^{\dagger n}_i}{\sqrt{n!}}
\label{compo}
\end{equation}
Then, the physical
Hilbert space is represented by the basis
\begin{equation}
|n, \{ n_i \} \rangle = \prod_i (c^\dagger_i)^{n_i} |0\rangle,
\label{C}
\end{equation}
where the product is taken over all lattice sites and $n_i$ can take only
two values $n_i=0,1$. As can be easily checked,  the annihilation and
creation operators of the composite fermions obey the standard
anti-commutation relation $[c_i,c^\dagger_j]_+=\delta_{ij}$ in the space
(\ref{C}).

The projection of the measured quantities into the space (\ref{C})
can be done by considering processes leading to
jumps of the composite fermions only.
If the composites are strongly bound and there are no free bosons or fermions,
this is a very accurate approximation. Then, we
ignore $b^\dagger_i b_j$ and $f^\dagger_if_j$ acting on $|n, \{ n_i \} \rangle$
unless $i=j$ which gives onsite local populations determined by presence
or absence of a composite on the site $i$.
So, in general
\begin{eqnarray}
b^\dagger_i b_j&=& nc^\dagger_ic_i\delta_{ij} ,
\label{b_com} \\
f^\dagger_i f_j &=&c^\dagger_ic_i\delta_{ij}.
\label{f_com}
\end{eqnarray}
Concerning the quartic operators $f^\dagger_if_jb^\dagger_k b_l$ in
eq.(\ref{I_2}) which are important for the detection of the FB
quasi-molecules, they must be set to zero (if acting on  $|1, \{ n_i \}
\rangle$) unless $i=j,\, k=l$ or $i=k, \, j=l$. For $n=1$, the following
relation  holds
\begin{eqnarray}
f^\dagger_if_jb^\dagger_k b_l =  \delta_{ik}\delta_{jl} (1-\delta_{ij})c^\dagger_i c_j +
\delta_{ij}\delta_{kl} c^\dagger_ic_ic^\dagger_kc_k.
\label{CC}
\end{eqnarray}
Similar relation can be established for any value of $n$ for the operator
linear in $f^\dagger_jf_i $ and of the order $n$ in $b_j^\dagger b_i$. We
will not, however, present it  explicitly here. Another set of useful
relations is:
\begin{eqnarray}
f^\dagger_if^\dagger_k f_lf_j &=&  [\delta_{ij}\delta_{kl}-\delta_{il}\delta_{kj}]c^\dagger_i c^\dagger_k c_kc_i,
\label{fff} \\
b^\dagger_ib^\dagger_kb_lb_j  &=&  n^2[\delta_{ij}\delta_{kl}+\delta_{il}\delta_{kj}]c^\dagger_i c^\dagger_k c_kc_i
\nonumber \\
&\,+&n(n-1)\delta_{il}\delta_{ij}\delta_{ik}c^\dagger_ic_i.
\label{bbb}
\end{eqnarray}

As long as the low energy properties are concerned in the low density limit,
the introduced truncation of the phase-space is accurate. Similar rules can
be formulated for the case of the F$\overline{\rm B}_n$ complexes.
In the case of $n>1$ the derivation of eq.(\ref{I_2_op}) proceeds as
follows. After substituting expressions (\ref{op_FF},\ref{op_BB}) into
eqs.(\ref{DF},\ref{DB}) and, then, into eq.(\ref{I_n+1}), we arrive at the
multiple lattice sum of the correlators in eq.(\ref{I_n+1}):
\begin{widetext}
\begin{eqnarray}
I_{n+1}({\bf Q})&\sim&\int d{\bf q}_2 ... d{\bf q}_{n+1}\sum_{i_1...i_{n+1}, j_1...j_{n+1}}{\rm e}^{i[({\bf Q}-{\bf q}_2 ... -{\bf q}_{n+1}){\bf x}_{i_1j_1} + {\bf q}_2{\bf x}_{i_2j_2} + ...+{\bf q}_{n+1}{\bf x}_{i_{n+1}j_{n+1}}]}
\nonumber \\
&\,&\langle \Psi|(f^\dagger_{i_1}f_{j_1} - \langle
\Psi|f^\dagger_{i_1}f_{j_1}|\Psi \rangle) :(b^\dagger_{i_2}b_{j_2} - \langle
\Psi|b^\dagger_{i_2}b_{j_2}|\Psi \rangle)...(b^\dagger_{i_{n+1}}b_{j_{n+1}}
- \langle \Psi|b^\dagger_{i_{n+1}}b_{j_{n+1}}|\Psi \rangle) :|\Psi \rangle
\label{croc}
\end{eqnarray}
\end{widetext}
It is enough to use the normal ordering $: ... :$ for bosonic operators only because just two fermionic operators
are involved.
 The terms with $i_1 \neq j_1$ which project into the one-particle density matrix of the composite fermions contain $2n$ bosonic operators together with the fermionic pair under $\langle \Psi| ... |\Psi \rangle$. These are
$\langle \Psi|f^\dagger_{i_1}f_{j_1}
:b^\dagger_{i_2}b_{j_2}...b^\dagger_{i_{n+1}}b_{j_{n+1}} :|\Psi \rangle =
\langle \Psi|f^\dagger_{i_1}(b^\dagger_{i_1})^n f_{j_1}(b_{j_1})^n |\Psi
\rangle
\delta_{i_1i_2}...\delta_{i_1i_{n+1}}\delta_{j_1j_2}...\delta_{j_1j_{n+1}}=
n!\langle \Psi|c^\dagger_{i_1}c_{j_1}|\Psi \rangle
\delta_{i_1i_2}...\delta_{i_1i_{n+1}}\delta_{j_1j_2}...\delta_{j_1j_{n+1}}$,
where the definition (\ref{compo}) is utilized.  This generates the first
term in the brackets of eq.(\ref{I_2_op}). The terms with $i_1=j_1$, after
being summed, produce zero due to the conservation of the total number of
fermions. This explains the presence of the last term in the brackets of
eq.(\ref{I_2_op}). Other terms carry less than $2n$ bosonic operators
averaged together with the fermionic ones. For example, one term which has
$2(n-1)$ bosonic operators is $\delta_{i_2j_2} \langle \Psi| b^\dagger_{i_2}
b_{i_2} |\Psi \rangle \langle \Psi|(f^\dagger_{i_1}f_{j_1} - \langle
\Psi|f^\dagger_{i_1}f_{j_1}|\Psi \rangle)
:b^\dagger_{i_3}b_{j_3}...b^\dagger_{i_{n+1}}b_{j_{n+1}} :|\Psi \rangle$,
where the factor $\delta_{i_2j_2}$ comes from the relation (\ref{b_com}). It
does not contribute to the terms with $i_1\neq j_1$ because the integration
over ${\bf q}_2$ in eqs.(\ref{I_n+1},\ref{croc}) has a structure $\int d{\bf
q}_2 \exp(i({\bf Q } -{\bf q}_2){\bf x}_{i_1j_1} + i{\bf q}_2{\bf
x}_{i_2j_2})\delta_{i_2j_2}$ and, thus, selects only terms with $i_1=j_1$.
The same logic applies to other terms with fewer number of bosonic operators.


\begin{references}
\bibitem{note2}
The composites F$\overline{\rm B}_n$ can be prepared from the bosonic Mott
Insulator state with the integer $k>n$ number of bosons per site. Then, each
fermion expels $n$ bosons from a site (which are removed from the system),
so that remaining number of bosons is $N_B= kN - nN_F$.
\end{references}
\end{document}